\documentstyle[emulateapj]{article}
\begin{document}

\title{Existence of Large Scale Synchrotron X-ray Jets in Radio-loud Active Galactic Nuclei}

\author{J.M. Bai\altaffilmark{1,2} and Myung Gyoon Lee\altaffilmark{1}}

\altaffiltext{1}{Astronomy Program, SEES, Seoul National University, Seoul 151-742, Korea
(jmmbai@astro.snu.ac.kr)}
\altaffiltext{2}{Yunnan Astronomical Observatory, NAO, the Chinese Academy of Sciences, 
Kunming 650011, China}

\begin{abstract}

In this paper, analytical arguments are presented that there exists a
synchrotron X-ray jet on large scales in most radio-loud AGNs,
based on the knowledge of the nature and physics of blazars.
In blue blazars and blue-blazar-like radio galaxies, the large scale X-ray 
jet may get faint along the jet, 
while in most red blazars and red-blazar-like radio galaxies, 
the X-ray jet is bright on 10 kpc scales
whether the jet is highly relativistic on large scales or not.
In extreme red blazars in which the jet is still highly relativistic on large scales and
the synchrotron peak of the inner jet lies in the infrared bands, 
the X-ray jet may get fainter along the jet 
from 10 kpc to 100 kpc scales while the optical and IR
jet gets brighter.
The predictions can be tested with the ongoing observations of 
the Chandra X-ray Observatory.

\end{abstract}


\section{INTRODUCTION}

Extragalactic jets are a spectacular feature associated with 
the activity of radio-loud active galactic nuclei (AGNs). 
They are pipelines through which the mass, momentum and energy are 
transported from the central nucleus to the extended radio lobes 
(e.g. Rees 1971; Begelman et al. 1984; Laing 1993), and
naturally play an important role for understanding the nature and 
physics of the invisible central engine in AGNs which is widely 
believed to be an accretion disk surrounding a super massive 
black hole (e.g. Rees 1984). 
Jets are usually observed in the radio bands with VLA on 
kiloparsec (kpc) scales and with VLBI on parsec scales. 
A handful of radio jets have been detected in the IR/optical bands 
by the Hubble Space Telescope on kpc scales.
At X-ray energies, 
about four nearby radio jets have been detected 
by EINSTEIN and ROSAT X-ray observatories in Cen A (Turner 1997; 
Feigelson et al. 1981), M87 (Neumann et al. 1997; Schreier et al. 1982), 
3C 273 (R\"oser et al. 2000; Harris and Stern 1987) and 
NGC 6251 (Mack et al. 1997), respectively. 

Since it was launched in July 1999, Chandra X-ray Observatory (CXO) has 
detected several more radio jets on kpc scales with high spatial and spectral resolution 
at X-ray energies.
While the radio (and probably the optical) emission of the jets is 
certain to be synchrotron, the radiation mechanisms responsible for the kpc scale 
X-ray jets have not been well understood yet. 
The X-ray jet in the distant quasar PKS 0637-752 has been explained in
synchrotron self-Compton (SSC) models (Schwartz et al. 2000) and
as inverse-Compton scattering of the cosmic microwave background 
(CMB, Tavecchio et al. 2000; Celotti et al. 2001). 
The X-ray jet in the famous quasar 3C 273 can be interpreted
as synchrotron emission, or synchrotron self-Compton emission
or inverse-Compton scattering of the CMB (Marshall et al. 2001; Sambruna et al. 2001; R\"oser et al. 2000).
The X-ray jet in M87, Cen A and Pictor A has also been interpreted in
several models (Neumann et al. 1997; Perlman et al. 2001; 
Turner et al. 1997; Wilson et al. 2001).
All these explanations are based on spectral fitting in the X-ray band or 
broad band spectra from radio to 
X-rays, uncertainty of which is usually high due to the weak signal of the jet. 

Blazars, including BL Lac objects and flat-spectrum radio quasars (FSRQs),   
are compact, flat spectrum radio sources
with highly variable and polarized nonthermal continuum emission extending 
up to X-ray and often gamma-ray energies, which are generally understood as
consequences of a relativistic jet oriented close to the line of sight 
(e.g. Blandford \& Rees 1978; Blandford \& K\"onigl 1979; Marscher 1980; 
Ghisellini et al. 1993; Kollgaard 1994).
The compact jet of a blazar which appears as a bright core in VLA maps but usually
exhibits an elongated jet in VLBI maps,
connects the central region with the outer,
more extended, large scale jets and radio lobes, and 
dominates the radio emission of the source, as in the cases of 3C 273 and 
PKS 0637$-$752. 
Blazars have been observed intensively from radio to gamma-rays, and the 
radiation mechanisms of the inner jets (the compact, parsec and subparsec scale jets) 
have been almost certain. 
The broad band spectral energy distributions (SEDs) of blazars have two
components, exhibiting a self-similar double-hump structure. 
Correlated variations across the SEDs are consistent with the picture
that a single electron population in the relativistic jet gives rise to 
both components, via synchrotron at low energies and
inverse-Compton scattering at high energies
(e.g. Ulrich et al. 1997; Ghisellini et al. 1998). 
In red blazars (such as 3C 273 and PKS 0637-752)
which have synchrotron peak at IR/optical wavelengths,
the X-rays are inverse-Compton emission, 
while in blue blazars which have synchrotron peak at UV/X-rays,
the X-rays are an extension of synchrotron emission
(Padovani \& Giommi 1996; Kubo et al. 1998). 
Besides the emission mechanisms, other aspects of the inner jets in blazars
have been intensively studied as well.

The knowledge about the inner jet in blazars may
serve as a starting point and a useful tool
to study the emission mechanisms in the kpc scale jet.
The kpc scale jet is an extension of the inner jet, connecting the inner jet to
the radio lobe, and thus must be somewhat similar to the inner jet.
Conclusions about the inner jet may be applied or generalized
to the kpc scale jet, and the conservative quantities can be used to
constrain the models for the kpc scale jet.
In this paper, we argue that there exists a synchrotron X-ray jet on kpc scales
in radio-loud AGNs, based on the knowledge of the inner jet in blazars.

\section{X-ray jets in blazars}

In blue blazars, the inner jet produces X-rays via 
synchrotron process. It is possible that the parsec scale synchrotron
X-ray jet extends to the kpc scale, 
i.e., there may be a synchrotron X-ray jet on kpc scales. 
In red blazars, the inner X-ray jet is inverse-Compton emission,
so it seems that red blazars cannot produce an X-ray jet on kpc scales
through synchrotron process. Furthermore,
for some sources, the trough
between the synchrotron and Compton peak happens to lie in the range 
of X-rays, and consequently there are not any X-ray jets in these sources
on kpc scales. However, it is probable that there exists a synchrotron X-ray jet 
in red blazars on kpc scales.
Statistical studies for complete samples of blazars revealed spectral
trends that as the bolometric luminosity increases, the luminosities
of the emission-lines and Compton component increase but the frequency of the
peak of the synchrotron component decreases (Sambruna et al. 1996; Fossati et al. 1998).
These trends support a simple paradigm in which the electron acceleration is 
similar in both red and blue blazars but the
relativistic electrons in the inner jet of red blazars suffer more Compton 
cooling because of larger external photon densities as indicated by the 
higher emission-line luminosity, leading naturally to lower characteristic
electron energies and thus lower frequencies of the synchrotron peak 
(Ghisellini et al. 1998; Urry 1999).
On kpc scales, the external photon densities of the jet in red blazars
are much smaller, so the electrons may reach much higher characteristic energies.
These high energy electrons may produce synchrotron X-ray jet 
on kpc scales, which can be seen below in detail.

\subsection{Physics of the inner jet in blazars}

Statistical analysis for all $\gamma$-ray loud blazars shows
a strong correlation between the characteristic electron energy
$\gamma_{peak}$ and energy density $(U_r+U_B)$ in the jet comoving frame 
for all $\gamma$-ray loud FSRQs (Ghisellini et al. 1998), 
\begin{equation}
\gamma_{peak} \propto (U_r+U_B)^{-0.5\pm0.06} = C(U_r+U_B)^{-0.5\pm0.06},
\end{equation}
where $U_r$ and $U_B = B^2/(8\pi)$ are the energy density 
of radiation (produced in or outside the jet) and magnetic field, respectively.
If including $\gamma$-ray loud BL Lac objects, 
the correlation is $\gamma_{peak} \propto (U_r+U_B)^{-0.6\pm0.04}$, slightly
different from Eq. (1). The difference may be caused by the
larger uncertainty of $U_r$, 
since BL Lac objects have weak or no emission lines, 
leading to larger uncertainties to the radiation energy density 
of the external radiation, $U_{ext}$.

The electrons must
be accelerated in the regions where they produce radiation
since the timescales of radiative energy losses of electrons are very
short to both synchrotron and Compton processes.
Assuming that the jet is composed of electrons and protons, and that 
the electron heating rate balances
to cooling rate $\dot{\gamma}_{heat} = \dot{\gamma}_{cool}$,
Eq. (1) can be deduced in the internal shock scenario (Ghisellini 1999),
suggesting that
$\gamma_{peak}$ is the result of the balance between heating and cooling.
Taking into account both synchrotron and inverse-Compton emission,
the electron cooling rate in the jet is 
$m_ec^2 \dot{\gamma} = (4/3)\sigma_Tc\gamma^2(U_{r}+U_{B})$, 
where $m_e$ is the mass of electron and $\sigma_T$ is the Thomson cross-section.
Using Eq. (1), at $\gamma = \gamma_{peak}$, 
$\dot{\gamma} \propto \gamma^2(U_{r}+U_{B}) \sim constant$, i.e.,
at $\gamma_{peak}$ the radiative cooling rate is nearly the same for all sources.
This may suggest that electron acceleration is the same in 
both red and blue blazars, independent of luminosity, $\gamma$ and U,
and that only the cooling differs (Ghisellini et al. 1998; Urry 1999). 

It is reasonable to assume that Eq. (1) can be applied to the kpc scale jet, 
and that the proportionality constant $C$ is the same on small and large scales.
The kpc scale jet is an extension of the inner jet.
It can be treated as an ``inner jet" originating
from a weaker ``nucleus", the end of the inner jet. 
Furthermore, on kpc scales the jet is still relativistic as indicated by 
the observed jet$-$counterjet intensity asymmetry.
The differences between the kpc scale jet and the inner jet are not more
than the differences between red and blue blazars. 
Probably, the electron acceleration mechanism in the kpc scale jet is 
the same as that in the inner jet, and the electron heating rate 
balances to the cooling rate. 

\subsection{The kpc scale synchrotron X-ray jet in blazars}

Given the relativistic electrons with $\gamma_{peak}$ in 
the magnetic field {\it B} of a relativistic jet with speed $\beta$ in  
units of the speed of light, Lorentz factor $\Gamma$ and 
viewing angle $\theta$, 
the peak frequency $\nu_s$ of synchrotron radiation in the observer's frame is  
\begin{equation}
\nu_s = 3.7\times 10^6\frac{\delta}{1+z}B\gamma_{peak}^{2},
\end{equation}
where $z$ is the redshift of the source and $\delta$ is the Doppler factor
$\delta = [\Gamma(1-\beta\cos\theta)]^{-1}$.
For blue blazars, $U_r(pc) \sim U_B(pc)$ in the inner jet. According to
Eq. (1), $\gamma_{peak}^2 = C^2[2U_B(pc)]^{-1} = 4C^2\pi / B_{pc}^2$. 
Substituting $\gamma_{peak}$ into Eq. (2) gives 
\begin{equation}
\nu_s(pc) = 3.7\times 10^6\frac{\delta_{pc}}{1+z}\frac{4\pi C^2}{B_{pc}}.
\end{equation}
On kpc scales, none of the jets shows any direct evidence of superluminal 
motion with one exception, M87 (Biretta et al. 1999), so in most cases the kpc scale jet
probably has only mildly relativistic speeds, i.e., $\Gamma(kpc)$ 
is slightly greater than 1. 
Since the inner jet is highly relativistic, 
the contribution of the emission from the inner jet to $U_r(kpc)$ cannot be neglected
(Celotti et al. 2001). The emission of the kpc scale jet itself also contributes a main part
of $U_r(kpc)$, but $U_r(kpc) \sim U_B(kpc)$ still holds.
According to Eq. (1),
$\gamma_{peak}^2 = C^2[2U_B(kpc)]^{-1} = 4C^2\pi / B_{kpc}^2$,
yielding
\begin{equation}
\nu_s(kpc) = 3.7\times 10^6\frac{\delta_{kpc}}{1+z}\frac{4\pi C^2}{B_{kpc}}.
\end{equation}
At point $r$ of the jet, the magnetic flux $L_B$ is
\begin{equation}
L_B = \pi\psi^2r^2c\Gamma^2U_B = \frac{c}{8}\psi^2r^2\Gamma^2B^2,
\end{equation}
assuming that the emitting region located at $r$ has a transverse 
dimension $R = \psi r$.
Conservation of magnetic flux $L_B$ gives
\begin{equation}
\frac{B(kpc)}{B(pc)} = \frac{\Gamma_{pc}}{1000\Gamma_{kpc}} \sim \frac{1}{100},
\end{equation}
i.e., $B(kpc) \sim 10^{-2}B(pc)$, assuming that $\psi$ is constant along the jet and 
$\Gamma_{pc} \sim 10$ (Blandford 1993; Marcher 1993; Celotti et al. 2001).
Combining Eqs. (3), (4) and (6) yields 
\begin{equation}
\nu_s(kpc) = 10^{2}\nu_s(pc)\frac{\delta_{kpc}}{\delta_{pc}} \sim 10 \nu_s(pc).
\end{equation}
The synchrotron emission of blue blazars peaks in the range of $\sim$0.01 keV to 
$\sim$1 keV, so
on kpc scales, the synchrotron peak of the jet emission lies above 0.1 keV, suggesting that
there exists a synchrotron X-ray jet in blue blazars on kpc scales.

For red blazars, $U_r / U_B$ is typically in the range of 10 -- 100, 
so $\gamma_{peak}^2 \sim 4 C^2\pi / (10 B_{pc}^2)$, yielding 
\begin{equation}
\nu_s(pc) = 3.7\times 10^6\frac{\delta_{pc}}{1+z}\frac{4\pi C^2}{10B_{pc}}.
\end{equation}
On kpc scales, as in the case of blue blazars, $U_r(kpc) \sim U_B(kpc)$, and
\begin{equation}
\nu_s(kpc) = 3.7\times 10^6\frac{\delta_{kpc}}{1+z}\frac{4\pi C^2}{B_{kpc}}.
\end{equation}
Thus 
\begin{equation}
\nu_s(kpc) = 10^{3}\nu_s(pc)\frac{\delta_{kpc}}{\delta_{pc}} \sim 10^{2}\nu_s(pc).
\end{equation}
The synchrotron peak of red blazars lies in the range of 
$\sim$10$^{13}$ Hz to $\sim$10$^{15}$ Hz,
so on kpc scales, the synchrotron peak of the jet emission lies 
above $\sim$0.01 keV, which is like those blue blazars
whose synchrotron peak lies in the UV and soft X-ray energies, 
suggesting that there exists a synchrotron X-ray jet
in red blazars on kpc scales as well.

\section{Discussion}

In Section 2, it has been argued that there exists a synchrotron 
X-ray jet in both blue and red blazars.
Blue and red blazars are just the extrema of the blazar class. Similarly,
there exists a synchrotron X-ray jet in intermediate blazars on kpc scales. 
Furthermore, 
according to the unified schemes, Fanaroff-Riley class I radio galaxies 
(Fanaroff \& Riley 1974) are intrinsically the same as BL Lac objects,
and Fanaroff-Riley class II radio galaxies and steep-spectrum 
radio quasars (SSRQs) are intrinsically the same as FSRQs, with the 
relativistic jet oriented at a larger
angle to the line of sight than blazars (e.g. Urry \& Padovani 1995). 
Therefore, radio galaxies and SSRQs have a synchrotron X-ray jet 
on kpc scales as well. 

According to Eqs. (4), (5) and (9), the synchrotron peak of the emission 
of the large scale jet shifts gradually to higher frequencies along the jet for 
both red and blue blazars. Consequently, in blue blazars and blue-blazar-like 
radio galaxies, the X-ray jet may get faint along the jet.
In typical red blazars and red-blazar-like radio galaxies, 
from $\sim$10 kpc to $\sim$100 kpc along the jet,
the synchrotron peak may lie rightly in the energy range of the CXO,
resulting in a relatively bright X-ray jet.
This may account for the X-rays observed from the ``inner jet" of 3C 273 
between 5$\arcsec$ and 10$\arcsec$ from the core (Marshall et al. 2001).
The viewing angle $\theta$ of the jet of 3C 273 to the line of sight is 
$\cos\theta = 0.95$ (Davis et al. 1991), so at the distance of 3C 273 ($z = 0.158$), 
 5$\arcsec$ and 10$\arcsec$ from the core 
of 3C 273 correspond to $\sim$38.43 kpc and $\sim$76.86 kpc from the core along the jet,
respectively, assuming $H_0 = 75$km$^{-1}$Mpc$^{-1}$ and $q_0 = 0.5$.
The X-rays from the ``inner jet" of 3C 273 are thus probably synchrotron emission.

If in some cases (possibly in some powerful FSRQs) the jet is still highly 
relativistic on large scales, i.e., $\Gamma \sim \Gamma_{pc} \sim 10$, 
the energy density of the CMB, $U_{CMB}$, gets significant
in the jet comoving frame. Assuming $L_B = 10^{45}$erg/s and redshift zero, 
 at $\sim$10 kpc $U_{CMB} \sim U_B$, and 
at $\sim$100 kpc $U_{CMB} \sim 100U_B$ (Celotti et al. 2001).
From $\sim$1 kpc to $\sim$10 kpc along the jet, the above analysis is still valid, 
except that Eqs. (7) and (10) turn to be 
$\nu_s(kpc) = 10^{3}\nu_s(pc)$ and $\nu_s(kpc) = 10^{4}\nu_s(pc)$, respectively.
On 100 kpc scales, 
\begin{equation}
\nu_s(100kpc) \sim 10^{-1}\nu_s(10kpc),
\end{equation}
indicating that the synchrotron peaks gradually shift to lower frequencies
from 10 kpc to 100 kpc scales. 
For extreme (and thus the most powerful) red blazars, 
in which the synchrotron peak of the inner jet lies in the infrared bands,
the synchrotron peak of the large scale jet may 
shift back to the optical band on 100 kpc scales, 
and consequently the X-ray jet gets fainter along the jet 
from 10 kpc to 100 kpc scales, while the optical and IR
jet gets brighter.
If in some powerful Fanaroff-Riley class II radio galaxies 
(counterparts of powerful red blazars) 
the jet is still highly relativistic on large 
scales, and is viewed at a very large angle,
Doppler beaming effects may cause the X-ray jet too weak to be detected.

\section{Conclusions}

In conclusion, we predict a detectable synchrotron X-ray jet on large 
scales in most radio-loud AGNs. 
In blue blazars and blue-blazar-like radio galaxies 
the large scale X-ray jet gets faint along the jet, 
while in typical red blazars and red-blazar-like radio galaxies  
the X-ray jet is bright on 10 kpc scales
whether the jet is highly relativistic on large scales or not.
In extreme red blazars in which the jet is still highly relativistic on large scales 
and the synchrotron peak of the inner jet lies in the infrared bands, 
the X-ray jet may get fainter along the jet 
from 10 kpc to 100 kpc scales while the optical and IR
jet gets brighter.

These predictions can be tested with the ongoing observations
of the Chandra X-ray Observatory. 
Although blazars are characterized by core-dominant morphology on
kpc scales, there are still a handful of blazars showing VLA jet. These are
blue blazars 0414+009, 548$-$322, 2201+044, 2155$-$304 
(Laurent-Muehleisen et al. 1993), 0829+089 and red blazars 
3C 371 (Wrobel \& Lind 1990), 
PKS 0521$-$365 (Keel 1986), 0954+658, 2007+777 (Kollgaard et al. 1992),
0752+258 (Antonucci \& Ulvestad 1985). 
In all these sources we predict a detectable synchrotron
X-ray jet on kpc scales.
Apart from some powerful Fanaroff-Riley class II radio galaxies in which
the jet may be highly relativistic on kpc scales 
($\Gamma_{kpc} \sim 10$) and oriented at a very large angle
to the line of sight, we also predict a detectable 
synchrotron X-ray jet on kpc scales in 
radio galaxies and steep-spectrum radio quasars.
As time goes on, more and more radio-loud AGNs will be observed by the CXO.
If a kpc scale X-ray jet is detected in each of these sources, 
the synchrotron origin of kpc scale X-ray jets will be verified.

We thank our referee, Dan Schwartz, for many suggestions which significantly
improved this paper. We also thank Gabriele Ghisellini 
for useful comments and suggestions. 
This work was financially supported by the BK21 Project
of the Korean government.

\end{document}